\documentclass[12pt,preprint]{aastex}

\shorttitle{Polarimetry of the Wolf-Rayet Binary CX Cep}
\shortauthors{Villar-Sbaffi et al.}

\begin{document}


\title{An Extreme Case of a Misaligned Highly Flattened Wind in the Wolf-Rayet Binary CX Cephei}

%% Note that \email has replaced the old \authoremail commandrk an email address

\author{A. Villar-Sbaffi\altaffilmark{1}, N. St-Louis\altaffilmark{1}, Anthony F. J. Moffat\altaffilmark{1} and Vilppu Piirola\altaffilmark{2,3}}

\altaffiltext{1}{D\'epartement de Physique, Universit\'e de Montr\'eal, C.P. 6128, Succursale Centre-Ville, Montr\'eal, QC H3C 3J7, Canada; and Observatoire du Mont M\'egantic.}

\altaffiltext{2}{Tuorla Observatory, University of Turku, FIN-21500, Piikkio, Finland.}

\altaffiltext{3}{Specola Vaticana, V-00120 Citta del Vaticano.}

\email{alfredo@astro.umontreal.ca, stlouis@astro.umontreal.ca, moffat@astro.umontreal.ca, piirola@utu.fi}


\begin{abstract}

CX Cep (WR 151) is the WR+O binary (WN5+O5V) with the second shortest period known in our Galaxy. To examine the circumstellar matter distribution and to better constraint the orbital parameters and mass-loss rate of the WR star, we obtained broadband and multi-band (i.e. UBVRI) linear polarization observations of the system. Our analysis of the phase-locked polarimetric modulation confirms the high orbital inclination of the system (i.e. $i=65^o$). Using the orbital solution of Lewis et al. (1993) we obtain masses of $33.9 M_{\odot}$ and $23.9 M_{\odot}$ for the O and WR stars respectively, which agree with their spectral types. A simple polarimetric model accounting for finite stellar size effects allowed us to derive a mass-loss rate for the WR star of $0.3-0.5\times10^{-5} M_{\odot}/yr$. This result was remarkably independent of the model's input parameters and favors an earlier spectral type for the WR component (i.e. WN4). Finally, using our multi-band observations, we fitted and subtracted from our data the interstellar polarization. The resulting constant intrinsic polarization of $3-4\%$ is misaligned in relation to the orbital plane (i.e. $\theta_{CIP}=26^o$ vs. $\Omega=75^o$) and is the highest intrinsic polarization ever observed for a WR star. This misalignment points towards a rotational (or magnetic) origin for the asymmetry and contradicts the most recent evolutionary models for massive stars (Meynet \& Maeder 2003) which predict spherically symmetric winds during the WR phase (i.e. $CIP=0\%$).

\end{abstract}


\keywords{Polarization -- binaries: eclipsing -- stars: Wolf-Rayet -- techniques: polarimetric -- stars: individual: \objectname{CX Cep} -- stars: mass loss}

\section{Introduction}

Wolf-Rayet stars have always puzzled astronomers in their attempt to explain their extreme mass-loss and dense winds. Binarity was initially believed to be the only mechanism capable of expelling their outer layers, expose their CNO-enriched cores and produce their strong winds. However, the existence of single Wolf-Rayet stars with similar characteristics as their double-star counterparts was a big drawback for this scenario.

Radiative (line-driven) wind acceleration (Cassinelli {\&} Castor 1973) is now the widely accepted explanation for the Wolf-Rayet phenomenon in massive stars. The strong UV radiation produced by massive stars coupled with significant metallicity provides a suitable environment for WR-type winds to develop. The inclusion of rotation in massive star evolutionary models is now able to reconcile theoretical predictions with previously unexplained observational evidence (Maeder {\&} Meynet 2000). For instance, the observed ratio of nitrogen-rich (WN-type) WR stars to carbon-rich (WC-type) WR stars is better reproduced by models that include rotation.

One important prediction of stellar evolution with rotation is that massive stars with solar or higher metallicities should reach the Wolf-Rayet phase with small rotational velocities even if their O-star progenitors rotate with speeds of up to 300 km/s (Meynet {\&} Maeder 2003). As massive stars age and lose their mass, most of their angular momentum is also carried away and stars with the strongest winds (i.e. in high metallicity environments) should be most affected by this spin-down process. Support for this prediction was provided by Harries et al. (1998) in their spectropolarimetric survey of a complete sample of northern WR stars. These authors found no evidence for a high frequency of highly asymmetric envelopes around WR stars, as would be the case for fast rotating stars (Maeder 1999). Their conclusions were based on the rarity of emission-line depolarization in the spectra of the WR stars surveyed. In fact, they found only 3 out of 16 stars showing evidence of line depolarization at the $1\sigma $ level.

In this paper, we report on our high-precision linear polarimetric observations of the Galactic WR binary CX Cep (WR 151) comprising an O5V star and a nitrogen-rich WN5 star (Lewis et al. 1993). CX Cep has the second shortest period of all known WR+O systems in our Galaxy (i.e. $P=2.12d$) with very shallow photometric eclipses suggesting a relatively small orbital inclination (i.e. Lipunova \& Cherepashchuk 1982 found $i\sim50^o$ based on the light-curve solution). Polarimetric observations by Kartasheva (2002b) and Schulte-Ladbeck {\&} van der Hucht (1989) favor, however, a much higher inclination (i.e. $i\sim70^o-80^o$ based on our reanalysis of their data; see Section 3). Kartasheva (2002b) also noted large epoch-to-epoch fluctuations in both the degree of matter concentration towards the orbital plane and the polarimetric amplitude, similar to those observed for the 1.64d WR binary CQ Cep (Villar-Sbaffi et al. 2005 and Kartasheva et al. 2000).

Our goals for this paper are to obtain a clearer picture of the matter distribution surrounding CX Cep using high-precision polarimetry with the highest time-resolution possible. In Section 2, we present our observations and method of reduction followed by our broadband and multi-band results in sections 3 and 4, respectively. In section 5, we discuss the results and draw our final conclusions.

\section{Observations and Reduction}

We used the 2.1m telescope of the McDonald observatory equipped with a rotating Glan-prism polarimeter for our broadband multi-epoch observations (see Breger 1979). To maximize the photon count-rate and because wavelength-independent Thomson scattering is the main polarizing mechanism in the winds from massive stars, our observations were carried out without a filter using a Hamamatsu R943A-02 photomultiplier with extended red response. Each integration lasted approximately 30 minutes and included observations of the moonlit sky to subtract background polarization. We chose our integration times in order to maintain the accuracy based on photon statistics\footnote{The accuracy based on photon statistics alone (i.e. Poisson statistics) should be regarded as a lower limit. Bastien (1982) showed that a more accurate estimate of the actual error can be found by multiplying this value by 1.5 in accordance with the spurious variability of standard polarized stars.} close to 0.03{\%} in both Stokes parameters Q and U.

Our five-band observations were carried-out at the 2.5m Nordic Optical Telescope in the Canary Islands using the TURPOL photopolarimeter (Piirola 1988) equipped with a rotating achromatic half-wave plate for linear polarimetry. This instrument allowed us to obtain simultaneous polarimetric observations in the UBVRI bands by using a combination of filters and dichroics resulting in effective wavelengths of 360nm, 440nm, 530nm, 690nm and 830nm for each band, respectively. The integration times were chosen such that the accuracy based on the least square-fit to the eight positions of the half-wave plate was less than 0.06{\%} in the B band and the total integration time was less than 60 minutes. Sky background polarization was automatically subtracted using a plane-parallel calcite plate as the polarizing beam splitter (Piirola 1988).

A summary of our observations is presented in Table 1.

For both sites we chose the standard polarized star HD 204827 (Hsu {\&} Breger 1982) to calibrate the position angle. This well-documented standard remained constant, within $\sim1^{o}$, throughout our runs and provided a reliable calibration. For our McDonald observations, HD 154345 (Gehrels 
1974), HD 9407 (Gliese 1969) and HD 212311 (Schmidt, Elston, {\&} Lupie 
1992) were used as standard unpolarized stars to determine the instrumental polarization. Numerous observations each night allowed us to determine an average value for the instrumental polarization in the Stokes parameters Q and U, which was then subtracted from our data.

According to the operating manual of the TURPOL polarimeter at the NOT, the instrumental polarization attributable to the optics can be considered negligible in all bands. This was confirmed to an accuracy (based on photon statistics) of $\sim0.03{\%}$ by our own observations of the unpolarized standards HD 154345 (Gehrels 1974), HD 9407 (Gliese 1969) and HD 67228 (Piirola 1977). Therefore, we did not subtract any instrumental polarization from our NOT observations. The details of all these calibration measurements are presented in Table 2.

Our observations were analyzed using the standard Brown, McLean, {\&} Emslie (1978; hereafter BME78) Fourier method, which assumes an optically-thin, corotating matter distribution with point sources, in order to determine the orbital parameters and the shape (i.e. moments) of the matter distribution. Accordingly, we represented the phase-locked variability of the Stokes parameters Q and U by fitting to our data (using $\chi^2$ minimization) a Fourier expansion up to second harmonics:

\[
\begin{array}{l}
 U\left( \phi \right) = U_0 + U_1 \cos \lambda + U_2 \sin \lambda + U_3 \cos 
2\lambda + U_4 \sin 2\lambda, \\ 
 Q\left( \phi \right) = Q_0 + Q_1 \cos \lambda + Q_2 \sin \lambda + Q_3 \cos 
2\lambda + Q_4 \sin 2\lambda. \\ 
 \end{array}
\]

Where $\lambda=2\pi\phi$ and the light-curve phase $\phi $ was calculated using the ephemeris for primary minimum (i.e. WR star in front at phase 0.0) of Kurochkin (1985):

\[
HJD_{\min } = 2444451.423 + 2.126897E.
\]

Although we expect WR binaries like CX Cep to increase their orbital period as they lose mass through stellar winds (see for instance the case of V444 Cyg in St-Louis et al. 1993), $\dot{P}$ in these systems is usually less than $0.1 sec/yr$ and can be neglected.

From the coefficients of this Fourier expansion, we calculated the orbital inclination $i$ and the orientation of the line of nodes $\Omega $ according to the BME78 prescriptions. Our definition for the orientation of the line of nodes comes from Harries {\&} Howarth (1996) in which $\Omega $ represents the angle of the ascending node measured from North through East with the constraint $\Omega<180^{o}$ (i.e. an ambiguity of $180^{o}$ exists in the determination of $\Omega $ since it is impossible from polarimetry to discern the ascending from the descending node). This definition differs from the one used by other authors, where $\Omega_{QU} $ represents the orientation of the major axis of the ellipse traced in the QU plane. Finally, the $\gamma_{i}$ moments of the matter distribution (see BME78) allowed us to calculate $\tau_{o}G$ and $\tau_{o}H$, describing respectively the degree of asymmetry about the orbital plane and the effective concentration of matter towards the orbital plane.

The errors on the parameters were determined using a Monte-Carlo method by adding Gaussian noise to our data using the Poisson error associated with each point multiplied by a factor 1.5\footnote{ We multiplied our Poisson errors by a factor 1.5 in order to comply with the findings of Bastien (1982).} as the standard deviation of the distribution. In this manner, we generated 1000 synthetic data sets and evaluated the orbital parameters for each set using the same Fourier method described above. We defined our error on a particular parameter as the interval for which 95{\%} (i.e. $2\sigma$ ) of our synthetic data sets could be found.

Because CX Cep has a small orbital separation (i.e. a$\sim25 R_{\odot}$; Lewis et al. 1993) and shows photometric eclipses, the system is expected to depart from the point-source approximation of BME78. Fox (1994) showed that a complete model for the polarimetric eclipses must take into account the reduction of direct unpolarized light during eclipses, the occultation of the shadowed regions behind the stars and the non-uniform illumination of electrons located close to the stars. Drissen et al. (1986) developed a first-order correction for the reduction of the direct unpolarized light during eclipses that they successfully used to correct for the phase-locked light-curve variability in the short-period WR binary CQ Cep (see also Villar-Sbaffi et al. 2005). However, because CX Cep's light-curve shows only small fluctuations in intensity (i.e. 0.13 and 0.04 magnitude dips at primary and secondary eclipses respectively), we neglected this correction in our analysis. A detailed discussion on the departures from the BME78 point-source approximation will be presented later in this paper.

\section{The Polarimetric Orbit}

The goals of our broadband (Texas) observations of CX Cep were to obtain the most precise polarimetric orbit possible and study the epoch-to-epoch and non-BME78 polarimetric variability of the system. Unfortunately, highly variable atmospheric conditions coupled with the presence of the moon on our last two missions (i.e. 2001 and 2002) affected the reliability of our polarimetric measurements of this relatively faint system (V=12.5 mag). This resulted in noisier, less-reliable, data sets with fewer measurements than our 2000 data. Our observations as a function of light-curve phase in the Stokes Q and U representation are shown in Figure 1.

\subsection{BME78 Analysis}

Using the BME78 diagnostics, we obtained the orbital parameters of CX Cep. The coefficients of our fit are presented in Table 3 along with the resulting orbital parameters. We also revisited the data of Kartasheva (2002b) and Schulte-Ladbeck {\&} van der Hucht (1989) and applied the same method of reduction that we used for our own data. The errors on the polarimetric measurements by Kartasheva (2002b) are quoted as being in the range 0.11{\%} to 0.15{\%} with no information regarding the errors on each single measurement. Therefore, for the purpose of this analysis, we assumed an error of 0.13{\%} on each individual measurement. For Schulte-Ladbeck {\&} van der Hucht (1989), we used the reprocessed data (corrected using the ephemeris of Kurochkin 1985 and averaged over the two channels) presented in Kartasheva (2002a) and the original Poisson errors. The resulting parameters are also presented in Table 3. 

We note that these revised results are considerably different from those published originally. Kartasheva (2002a and 2002b) probably used a different set of errors (i.e. weights) from the ones assumed in this paper to perform the $\chi^{2}$ minimization. This strong instability in the solutions is reflected by the high errors on the resulting orbital parameters. We also note that the large differences in the $Q_0$ and $U_0$ constants between the Kartasheva (2002b) V-band observations and the other unfiltered (i.e. white-light) observations is easily explained by the wavelength-dependence of the high interstellar polarization component (see Section 4) which usually peaks in the V-band.

Our multi-epoch results for the orbital parameters agree (within $2\sigma$ ) with each other and with the (revised) results from other authors, although our last two years are plagued by higher uncertainties due to the noisier data sets. Because the system had a more symmetric matter distribution about the orbital plane compared to other epochs (i.e. $\tau_{o}G \ll \tau_{o}H$), we believe that our 2000 data provide a better representation of the orbit and matter distribution around CX Cep within the BME78 assumptions.

\subsection{Phase-Locked Non-BME78 Variability}

Phase-locked non-BME78 variability in polarimetric data has already been found in a few WR binaries (e.g. St-Louis et al. 1993 and Villar-Sbaffi et al. 2005). In short-period binaries, the most important cause of BME78 departures is the violation of the point-source assumption due to the comparable size of the orbital separation and the stellar radii (e.g. $a\sim 25 R_{\odot}$ and $R_{O}+R_{WR}\sim 15 R_{\odot}$ for CX Cep). 

In order to better appreciate these deviations from the predicted double-wave behavior in CX Cep, we binned our 2000 data (i.e. the most reliable data set) in phase intervals of 0.03. This binning allowed us to average-out the stochastic polarimetric fluctuations caused by the presence of fast-moving blobs in the wind of the WR star (see Robert et al. 1989)\footnote{Here we assume that the spatial distribution of blobs over a long period of time (i.e. 2 weeks) follows the density distribution of the wind.}. The effect of these blobs on the polarization should be maximal around phases 0.0 and 0.5 due to the alignment along the line-of-sight of the stars during eclipses. This is in agreement with the larger scatter observed in our data around those phases. Finally, we rotated the data by $-\Omega_{QU}=30^{o}$ in the QU plane\footnote{ This value was chosen from our most reliable 2000 data and is related to the orientation for the line of nodes $\Omega $ by the relation $\Omega _{QU} = 2\Omega - 180$.} to the orbital plane of symmetry of the system and plotted the polarization in Stokes Q and U as a function of light-curve phase (Figure 2). A clear non-BME78 (i.e. non-double-wave) structure appears around phase 0.5 when the O star eclipses the WR star. A similar structure has been observed in the WR binaries V444 Cyg (St-Louis et al. 1993) and CQ Cep (Villar-Sbaffi et al. 2005). This non-BME78 modulation was successfully modeled for V444 Cyg by considering the effect of the gradual eclipse by the O star of the dense inner regions of the WR wind during secondary eclipse. This simple model allowed the authors to constrain the stellar radii of both stars along with the mass-loss rate and velocity law of the WR star.

Following this success, we decided to develop a similar model to simulate the polarimetric variability in CX Cep not only at the eclipses but over the entire double-wave orbit. Our main goal is to determine the mass-loss rate of the WR star. We used the same formalism as St-Louis et al. (1993) and solved the modified BME78 integrals that assume an optically-thin wind corotating in the frame of the binary. The finite size of the stars was accounted for by setting the electron density equal to zero in the shadowed regions behind the stars and correcting for the depolarization of photons scattered close to the stars using the Cassinelli et al. (1987) factor.

The orbital separation of $25 R_{\odot}$ from Lewis et al. (1993) was used and the magnitude difference $M_{WR}-M_{O}=1.4$ was taken from van der Hucht (2001). We assumed that the WR wind consists exclusively of ionized helium and followed the prescriptions of Moffat {\&} Marchenko (1993) to calculate the number of free electrons per nucleon in the wind for a typical WNE star. The radius of the O star was kept fixed at 11 $R_{\odot}$ based on the most recent models of Martins et al. (2005) for an O5V star. Although the spectral type of the O star companion is still a source of debate (Kartasheva 2002a), our models were not very sensitive to changes in the O-star radius of $\pm2R_{\odot}$ (see Table 4). 

We initially fixed the orbital inclination of our models at $73^o$ based on the 2000 broadband results. However, we quickly realized that a lower inclination produced a much better fit to the binned data. An inclination of $65^o$ was finally adopted. This value differs by $4\sigma$ from the one determined by the BME78 analysis. However, this discrepancy agrees with the results of Fox (1994) who showed that by considering occultation effects in the envelope of a modeled binary system, deviations of $\sim3^{o}$ could be found in the inclination derived by Fourier analysis of a synthetic polarimetric orbit.

The integrals were solved within a volume equal to 10 times the orbital separation by Monte-Carlo integration using a Sobol quasi-random number sequence in three dimensions with a word-length of 30 bits. The major improvement of our model was to replace the cut-off radius used by St-Louis et al. (1993) by the optical-depth $\tau $ along the line-of sight, therefore replacing an arbitrary parameter (i.e. the cut-off radius) by a physical one. The cut-off radius was initially introduced in order to exclude high density regions in the inner wind where multiple-scattering leads to significant depolarization.

The optical-depth for a scattered photon at coordinates $(x_{o},y_{o},z_{o})$ was calculated by assuming that the only source of opacity was Thomson scattering:

\[
\tau = \int\limits_{x_o }^\infty {n_e \left( {x,y_o ,z_o } \right)\sigma _T 
dx}. 
\]

Where $\sigma _{T}=6.65\times10^{ - 25} cm^{ - 2}$ is the Thomson scattering cross-section. The electron density $n_{e}(x,y,z)$ was determined using the principle of mass conservation at every point in the accelerated wind using a $\beta$-type velocity law with $\beta=5$ and a sonic radius (i.e. $R_s$) of $2 R_{\odot}$ for a typical WNE star. The high value of $\beta$ and the small sonic radius of the wind follow the optically-thick radiation-driven wind models of Nugis \& Lamers (2002) who were able to reproduce the high mass-loss rates of Wolf-Rayet stars\footnote{A lower $\beta$-value with a larger sonic radius (i.e. $\beta=1$ and $R_s=4 R_{\odot}$) was also found to fit our data without affecting considerably the determined mass-loss rate.}. $\tau_{max}=2/3$ was chosen as the maximum optical-depth beyond which a photon became unpolarized and did not contribute to the total polarization. This optical depth corresponds to the adopted definition of a star's photosphere and resulted in a good fit to the data (see models M10 and M11 in Table 4). However, because we assume that Thomson scattering is the only source of opacity and neglect line transitions, we are in fact  underestimating the photospheric radius and overestimating its Thomson scattering optical-depth (i.e. $\tau_{max}$). As our models will prove (see Table 4), a lower $\tau_{max}$ will result in a higher mass-loss rate for the WR star. Therefore, the mass-loss rate determined with $\tau_{max}=2/3$ represent a lower limit compared to the actual mass-loss rate.

In Figure 2 (solid line) and Table 4 (model M1) we present the best fitting model to our binned data. The model was fitted by $\chi^{2}$ minimization by first determining the horizontal (i.e. $\Delta\phi$) and vertical (i.e. $\Delta Q$ and $\Delta U$) off-sets and finally by fitting the value $\gamma=\dot{M}/V_{\infty}$ which controls simultaneously the amplitudes in Stokes Q and U (see St-Louis et al. 1993). Although our best fit is quite good, there are still a few structures in Stokes Q at phases 0.0 and 0.5 that could not be reproduced. These structures are probably signs of a complex matter distribution which might include a distorted wind-wind collision zone and a disk-like density enhancement of the wind (see section 4.2). To avoid any problems caused by these structures on our final results we excluded from the $\chi^{2}$ minimization any data points in Stokes Q located within $\pm0.1$ in phase around the eclipses.

Assuming a characteristic terminal velocity for a WN5 wind of 1700 km/s, we obtain a mass-loss rate for the WR component of $0.3\times10^{-5} M_{\odot}/yr$. A few other models showing the effect of varying the orbital inclination (models M2 and M3), the exponent $\beta$ of the velocity law (models M4 and M5), the star's radii (models M6, M7, M8 and M9) and the maximum optical depth (models M10 and M11) are also shown in Table 4. Surprisingly, the mass-loss rate determined from all the models is remarkably stable and only fluctuates by about $0.2\times10^{-5} M_{\odot}/yr$ around the value determined from our best fitting model. As we stated above, the adopted maximum Thomson scattering optical depth (i.e. $\tau_{max}=2/3$) is our main source of uncertainty and, in fact, results in a lower limit for the mass-loss rate of the WR star. However, because spectral lines only amount to approximately 10\% of the total flux from a WR star, we do not believe that $\tau_{max}$ could be considerably lower than our assumed value of $2/3$.

We also note that a small bias towards low mass-loss rates is inherent to our polarimetric model since we neglect the weaker O-star wind. In relation to WR stars, O-stars have lower mass-loss rates and faster winds and, therefore, lower densities in their circumstellar matter distribution. Since the fitted parameter $\gamma$ (i.e. $\gamma = \dot{M}/V_{\infty}$) from which we derive the mass-loss rate is actually an average over the whole wind, the low-$\gamma$ contribution from the O-star wind will decrease the mass-loss rate determined from our model. Fortunately, the matter distribution surrounding CX Cep is dominated by the strong WR wind (see Lewis et al. 1993, their Figure 14) and this bias is probably not very high.

Finally, based on the resulting mass-loss rates from all our models, we believe that the mass-loss rate of the WR star is in the range $0.3-0.5\times10^{-5} M_{\odot}/yr$, with a preference towards the higher value. A detailed two-wind model including wind-wind collision effects would certainly provide a more accurate result.

\section{The Constant Intrinsic Polarization}
\label{sec:mylabel1}

By measuring the constant intrinsic polarization (CIP) of a star, we are in fact measuring the degree of spherical asymmetry of the matter distribution surrounding it. A high CIP implies a strong departure from spherical symmetry. In order to determine the CIP of a star we must fit and remove the contribution of the interstellar polarization to the total observed polarization. For massive stars, where (wavelength-independent) Thomson scattering is the dominant source of linear polarization, this is usually done by obtaining multi-wavelength (i.e. UBVRI) polarimetric observations of the star in order to subtract the wavelength-dependent interstellar contribution. Because in this section we are only interested in the CIP of the system, our multi-band observations are scarcer in phase coverage and serve essentially to determine the zero-point of the polarimetric variability in each band (see Figure 3).

We note that a small wavelength-dependent intrinsic contribution to the polarization is expected due to the presence of strong emission lines in the spectrum of a WR star (see for instance Moffat \& Piirola 1993). These lines form far from the dense central regions of the wind and their polarization can differ from that of the surrounding continuum (see Villar-Sbaffi et al. 2005). However, the line-flux in a standard Johnson band only represents $\sim10\%$ of the total flux from the star and we decided to neglect this effect in our analysis. The results of the BME78 analysis of the multi-band observations are presented in Table 5.

The CIP of CX Cep was determined using the method described in Villar-Sbaffi et al. (2005). We assumed that the constant interstellar polarization could be represented by a modified Serkowski's law (Serkowski, Mathewson, {\&} Ford 1975):

\[
\begin{array}{l}
 U_{IS} (\lambda ) = P_{\max } \sin \left[ {2\theta _{IS} } \right]e^{ - 
K\left( {\lambda _{\max } } \right)\ln ^2\left( {\frac{\lambda _{\max } 
}{\lambda }} \right)}, \\ 
 Q_{IS} (\lambda ) = P_{\max } \cos \left[ {2\theta _{IS} } \right]e^{ - 
K\left( {\lambda _{\max } } \right)\ln ^2\left( {\frac{\lambda _{\max } 
}{\lambda }} \right)}. \\ 
 \end{array}
\]

Where P$_{max}$ is the maximum interstellar polarization found at wavelength $\lambda _{max}$, $K\left( {\lambda _{\max } } \right) = 0.01 + 1.66\lambda _{\max } $ (Whittet et al. 1992) and $\theta _{IS} \left( \lambda \right) = C_1 + C_2/\lambda$ (Dolan {\&} Tapia 1986, Moffat {\&} Piirola 1993 and Matsumura et al. 2003). The constant Stokes parameters $Q_{0}$ and $U_{0}$ from the BME78 analysis can then be expressed as:

\[
\begin{array}{l}
 U_0 \left( \lambda \right) = U_{WR} - \frac{U_3 \sin ^2i}{1 + \cos ^2i} + 
U_{IS} \left( \lambda \right), \\ 
 Q_0 \left( \lambda \right) = Q_{WR} - \frac{Q_3 \sin ^2i}{1 + \cos ^2i} + 
Q_{IS} \left( \lambda \right). \\
 \end{array}
\]

Where $U_{WR}$ and $Q_{WR}$ are the components of the CIP vector and $U_{3}$, $Q_{3}$ and $i$ were taken from our fit to the broadband 2000 data (see Table 3).

Using these equations and a non-linear Levenberg-Marquardt code, we fitted simultaneously our values of $Q_{o}$ and $U_{o}$ to obtain the CIP of the system (i.e. $Q_{WR}$ and $U_{WR})$ and the parameters describing the interstellar polarization (i.e. $P_{max}$, $\lambda _{max}$, $C_{1}$ and $C_2$). The numerical weight of each band was taken to be the $rms$ error on the double-wave fit to Q($\phi$) and U($\phi$) found in Table 5 and not the actual errors on Q$_{o}$ and U$_{o}$ determined from the fit. We justify this statement by the fact that the constants Q$_{o}$ and U$_{o}$ only represent the sum of the interstellar and intrinsic polarizations to the extent that the BME78 assumptions hold for this system. Therefore, we believe that the $rms$ errors of the double-wave fit represent more accurately the relative weight to be given to each band. The resulting parameters for this first model (i.e. Model 1) are presented in Table 6 and the fitted curve along with our data can be found in Figure 4.

Because of the possible instability of our solution due to the weak $\lambda $-dependency of the interstellar polarization, we performed another fit with the constraint $Q_{WR}=U_{WR}=0{\%}$ to test the possibility that our first solution was statistically insignificant. The parameters of this second fit and the corresponding curve (i.e. Model 2) can also be found in Table 6 and Figure 4.

With an rms error less than half that of Model 2, Model 1 provides a much better fit to the constant polarization of CX Cep\footnote{The rms error presented in this paper is calculated using the formula: $rms = \sqrt{\sum\limits_{i = 1}^N {(y_i - y_{fit}(x_i))^2}/(N - M)}$, where M is the number of fitted parameters. This rms error gives an unbiased measure for the quality of the fit, which takes into account the reduced number of free parameters in Model 1 compared to Model 2.}. We can therefore say with certainty that the $CIP$ in not zero and probably very high. However, the resulting interstellar polarization seems very high. According to Serkowski, Mathewson {\&} Ford (1975), the maximum interstellar polarization observed for completely aligned dust grains in a purely regular external magnetic field is related to the extinction E(B-V) along the line-of-sight by the equation $P_{max}=9E(B-V)$. Using the absorption $A_{V}=3.77$ mag in van der Hucht (2001) and the commonly accepted ratio $R_{V}=3.1$ between the extinction and absorption in the V band we obtain a maximum interstellar polarization along the line-of-sight towards CX Cep of almost 11{\%}. This is larger than the value determined from Model 1 and, therefore, in accordance with our results although such a high polarization would necessitate an extreme case of dust-grain alignment. Another source of supporting evidence for this high interstellar polarization is the presence of a normal (i.e. should not show intrinsic polarization) B0.5V star within $0.11^{o}$ of CX Cep at a distance of $2.3$kpc (cf. $\sim6$kpc for CX Cep; van der Hucht 1988) with an average interstellar polarization of 4.5\% oriented at an angle of $41^{o}$ (Heiles 2000). This angle is the same as that determined from Model 1 but differs by $10^{o}$ from the interstellar polarization angle of Model 2, therefore supporting the validity of Model 1. 

The main source of uncertainty on our CIP determination is probably the scarce data coverage centered mainly on phases 0.0 and 0.5 where the scatter is larger (see Figure 1) due to the presence of blobs in the wind and the alignment of the light-sources during eclipses. However, since our multi-band observations cover more than three orbits (i.e. 8 nights) and the eclipses were therefore observed at least three times, we believe that any epoch-to-epoch fluctuations were averaged-out (i.e. assuming again that the blob distribution follows that of the wind over a long period of time). An estimate of the error caused by the stochastic polarimetric fluctuations can be obtained through the covariance matrix produced by the Levenberg-Marquardt method used to fit and remove the interstellar polarization. The diagonal elements of this matrix are the squared uncertainties on the fitted parameters. These uncertainties depend on the weights (i.e. errors) given to the data points that we assumed to be the rms errors on the BME78 double-wave fits. These rms errors measure the scatter in Q and U caused by blobs (and other non-BME78 behavior) around the mean BME78 curve; thus, they provide an estimate of the maximum vertical shift (i.e. in Stokes Q and U) that blobs can impart to our data. Therefore, the covariance matrix provides an upper-limit estimate of the error on our fitted CIP. Using this method we obtain a maximum error on the intrinsic polarization of $\sigma_{ P_{WR}}=1\%$, which is almost four times lower than the observed CIP (i.e. $P=3.89\%$).

Based on these facts, it appears that CX Cep has a very high intrinsic polarization (i.e. $P_{WR}\sim 3-4{\%}$) aligned at $26^{o}$ from the equatorial plane of reference. This angle doesn't correspond to either the orientation of the orbital plane (i.e. $\Omega$) or the polar axis (i.e. $\Omega+90$) of the system. A similar phenomenon was recently observed in the short-period WR binary CQ Cep (Villar-Sbaffi et al. 2005) and indicates an apparent lack of correlation between binarity and wind geometry in WR+O systems. To confirm this assertion, we performed a fit on the constants $Q_0$ and $U_0$ with the constraint $\theta_{WR}=\Omega+90^o$, as would be the case for a tidally distorted wind. The results of this third model have an rms error more than twice that of model 1 and confirm the misalignment of the CIP in relation to the orbital axis (see Table 6).

\section{Discussion and Conclusions}

Although it has the second shortest period of all know WR+O binaries in our Galaxy, CX Cep is a relatively poorly studied system. In this paper, we have confirmed the polarimetric orbital parameters determined by other authors and considerably improved their accuracy. The BME78-derived orbital inclination confirmed the almost edge-on configuration of the system and provided an accurate orientation for the line of nodes [i.e. $i = (73 \pm 2)^o$ and $\Omega =(75 \pm 1)^o$]. However, by including the finite stellar size effects in a model of the polarimetric orbit, we found a lower and more accurate inclination of $65^{o}$. Assuming an error on the orbital inclination of $3^o$, the resulting masses for both components can then be found using the orbital solution of Lewis et al. (1993):

\[
\begin{array}{l}
 M_O = \frac{25.2\pm 1.9\,M_ \odot }{\sin ^3i} = 33.9\pm 3.6\,\,M_ \odot \\ 
 M_{WR} = \frac{17.8\pm 1.4\,M_ \odot }{\sin ^3i} = 23.9\pm 2.6\,\,M_ \odot 
 \end{array}
\]

The mass of the O-star agrees with the spectroscopic masses for Galactic O5-6V stars (Martins et al. 2005) and the WR mass falls within the acceptable range for a rotating WNE star with an initial mass of $60R_{\odot}$ (Meynet \& Maeder 2003).

Our polarimetric model also allowed us to obtain the mass-loss rate of the WR component, which was found to lie in the range $0.3-0.5\times10^{-5} M_{\odot}/yr$ and was remarkably independent of the adopted input parameters. However, this value is quite low compared to other WN5 stars with mass-loss rates close to $1\times10^{-5} M_{\odot}/yr$. An earlier (i.e. hotter) spectral type (i.e. WN3-4) would be more appropriate for such a low $\dot{M}$ (Nugis \& Lamers 2002). In fact, Lewis et al. (1993) concluded that a WN4 spectral type was also acceptable for the WR component in CX Cep.

Our most surprising discovery is the presence of a large 3-4\% CIP in CX Cep, higher than that of the most polarized Be stars (Poeckert et al. 1979). Although the uncertainty on this value (and the corresponding high interstellar polarization) is probably high (i.e. up to 1\% in the worst case scenario), our analysis shows that the probability that the wind is spherically symmetric is very low. Based on the radiative transfer model presented in Harries et al. (1998), this CIP implies an equator-to-pole density ratio higher than 5. Like for the WR binary CQ Cep (Villar-Sbaffi et al. 2005), this departure from spherical symmetry is characterized by a CIP vector misaligned in relation to the natural axes of symmetry of a binary (i.e. the polar and equatorial axes) which points towards a rotational origin for the asymmetry. Alternatively, this could reveal the presence of a magnetic field which is affecting the symmetry of the wind.

Within the observations of Harries et al. (1998) and the models of Meynet {\&} Maeder (2003), CX Cep is an anomaly. Massive stars at solar metallicity should reach the WR phase with low rotational velocities and, therefore, spherically symmetric winds. For this reason, the extremely flattened wind of CX Cep should provide valuable information regarding the effects of rotation on massive star evolution and should be studied further.

\acknowledgments
AVS, NSL and AFJM are grateful for financial support from NSERC (Canada) and FQRNT (Qu\'ebec). We would also like to thank Andrei Berdyugin for his help operating the NOT telescope and for sharing his observing time with us.

\clearpage


\begin{figure}
\plotone{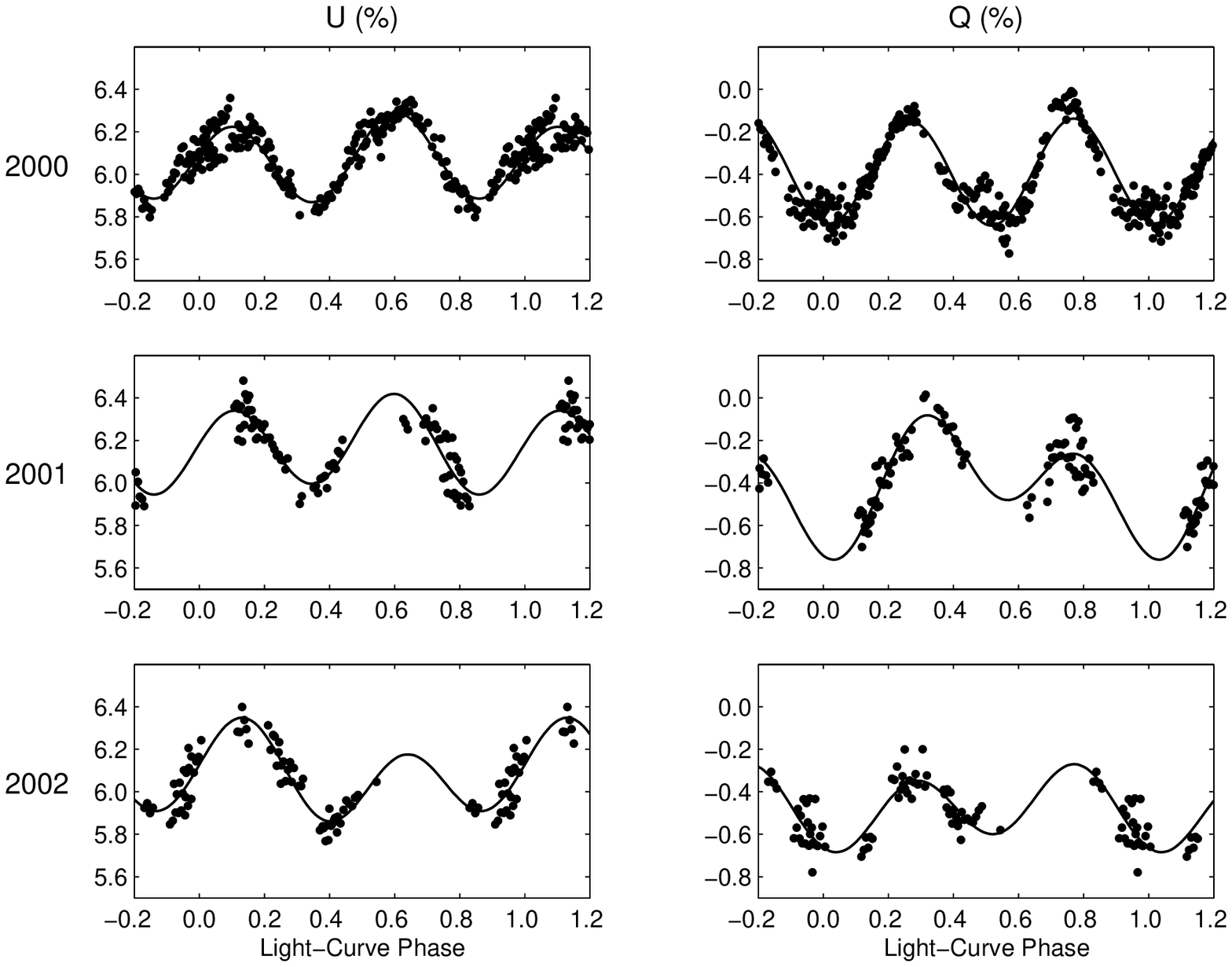}
\caption{Normalized $Q$ and $U$ Stokes parameters for our unfiltered McDonald observations as a function of the 2.12d light-curve phase of CX Cep. These observations were fitted separately for each epoch to a Fourier expansion up to second harmonics (solid line) following the prescriptions of BME78.}
\end{figure}

\begin{figure}
\plotone{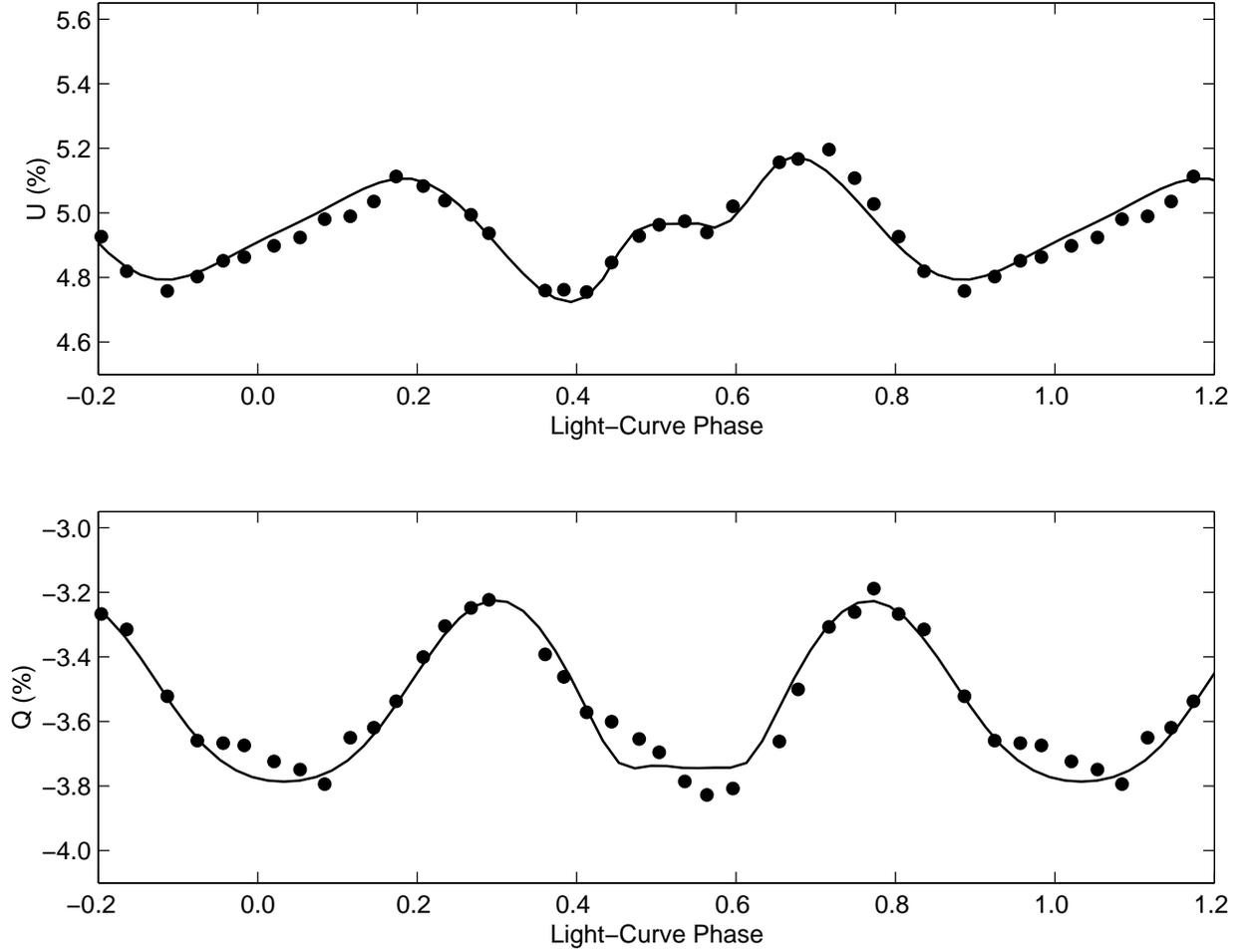}
\caption{Binned broadband observations of CX Cep as a function of light-curve phase rotated by $-\Omega_{QU}$ to the orbital plane of symmetry. The solid line represents our best fitting model (i.e. model M1 from Table 4) accounting for the polarimetric eclipse of the WR wind. }
\end{figure}

\begin{figure}
\plotone{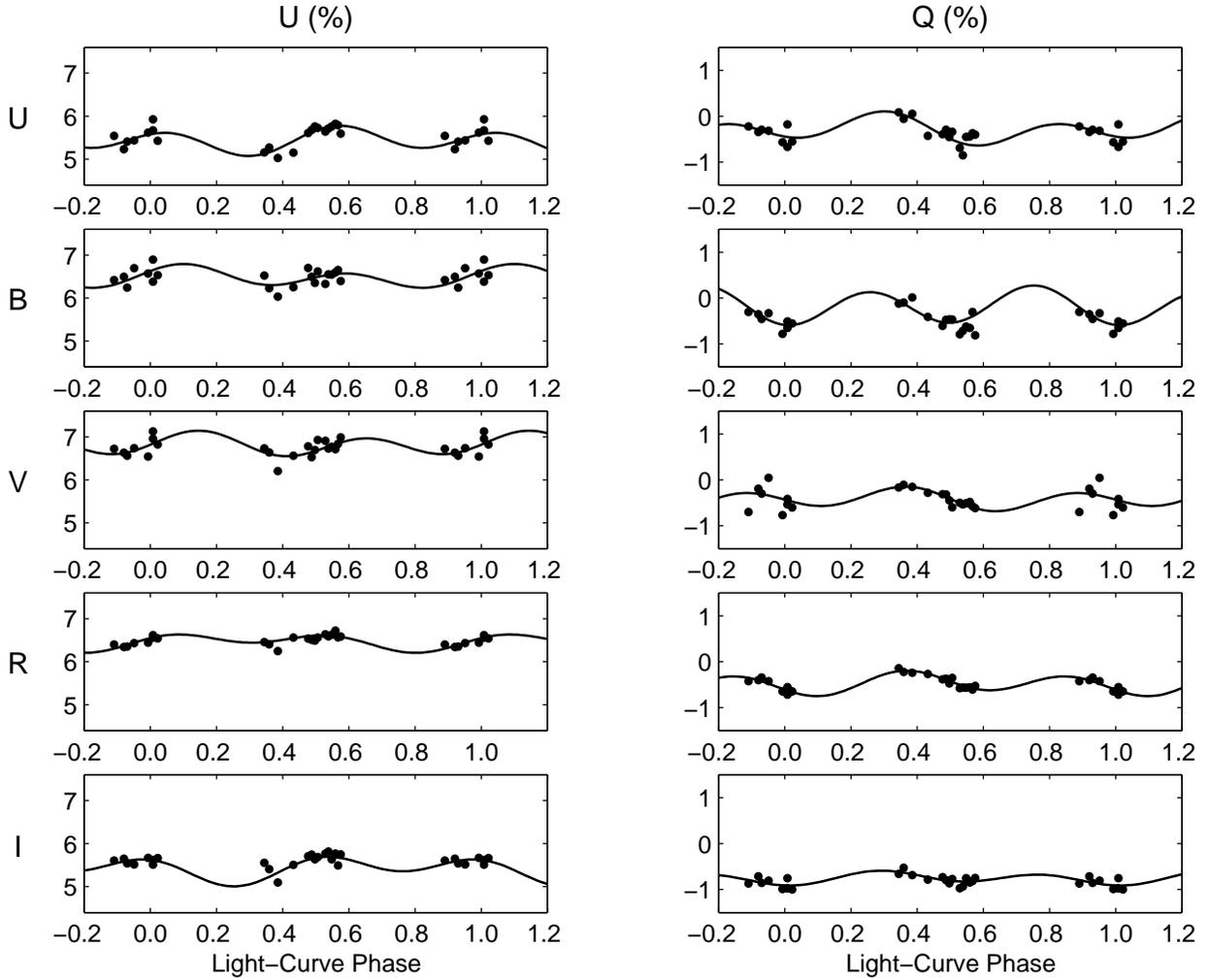}
\caption{Normalized $Q$ and $U$ Stokes parameters from our multi-band (UBVRI) NOT observations versus the 2.12d light-curve phase of CX Cep. These observations were fitted to a Fourier expansion up to second harmonics (solid line) following the prescriptions of BME78.}
\end{figure}

\begin{figure}
\plotone{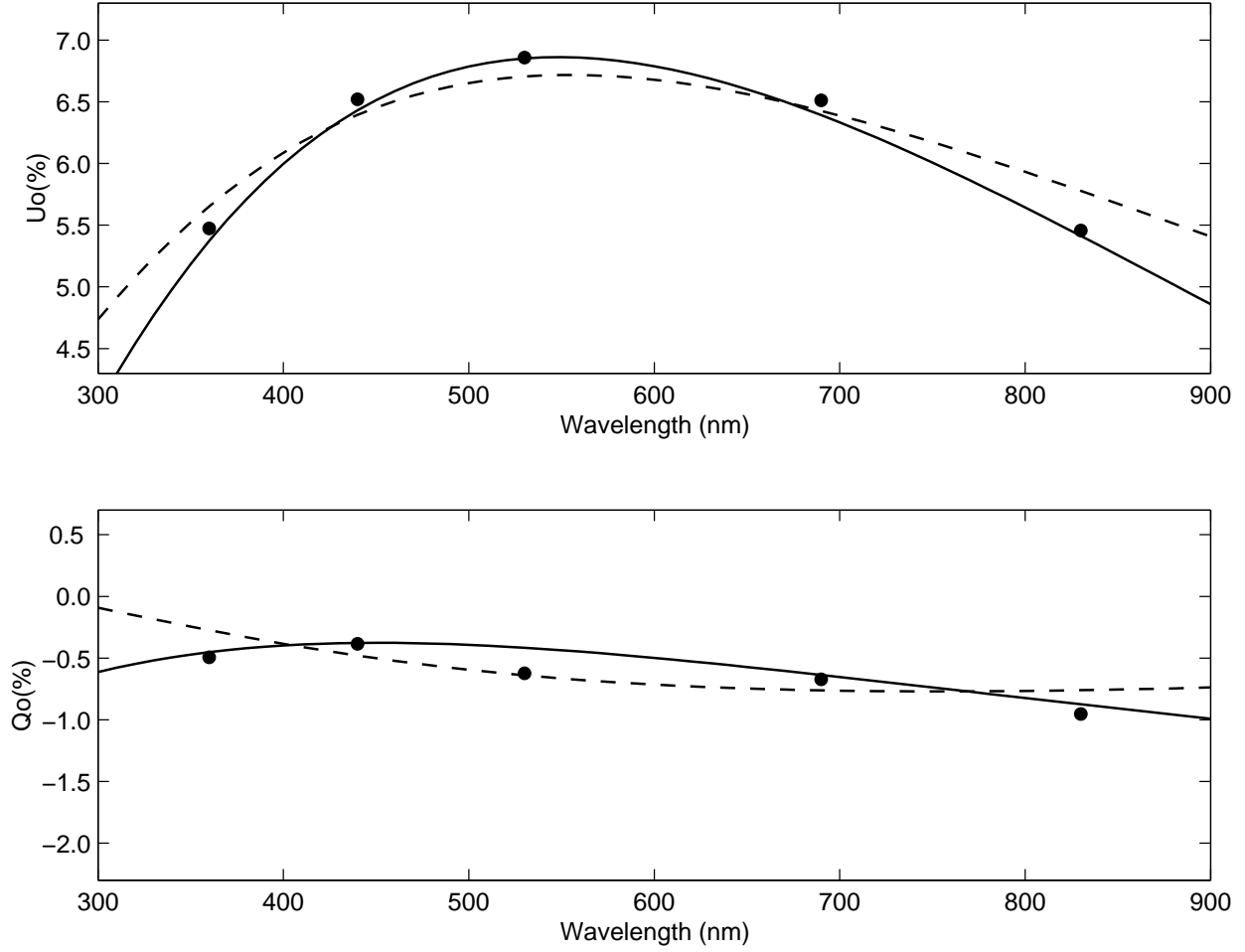}
\caption{Constant components of the polarization $Q_o$ and $U_o$ as a function wavelength obtained by fitting our NOT observations with a Fourier function up to second harmonics. The solid line represents our fit using a modified Serkowski function plus an additional wavelength-independent term corresponding to the intrinsic polarization of the system. The dashed line represents the same fit obtained with the constraint $Q_{WR}=U_{WR}=0{\%}$ (i.e. for a spherically symmetric wind). }
\end{figure}

\clearpage

\begin{deluxetable}{lcccc} 
\tablecolumns{5} 
\tablewidth{0pc} 
\tabletypesize{\footnotesize}
\tablecaption{Summary of Observations} 
\tablehead
{ 
\colhead{} 			& 
\multicolumn{2}{c}{Epoch 1 (2000)} &
\colhead{Epoch 2 (2001)} &
\colhead{Epoch 3 (2002)}
}

\startdata 

Observatory& 
McDonald& 
NOT& 
McDonald& 
McDonald \\

Telescope size& 
2.1m& 
2.5m& 
2.1m& 
2.1m \\

Instrument& 
Breger polarimeter& 
TURPOL& 
Breger polarimeter& 
Breger polarimeter \\

Filters& 
None& 
UBVRI& 
None& 
None \\

First night (UT)& 
August 31$^{st}$& 
October 8$^{th}$& 
August 30$^{th}$& 
September 10$^{th}$ \\

Number of nights& 
14& 
20& 
14& 
14 \\

\enddata 
\end{deluxetable}

\clearpage

\begin{deluxetable}{lccccccccc} 
\tablecolumns{10} 
\tablewidth{0pc} 
\tabletypesize{\footnotesize}
\tablecaption{Calibration Measurements} 
\tablehead
{ 
\colhead{} 			& 
\multicolumn{3}{c}{McDonald} &
\colhead{} &
\multicolumn{5}{c}{NOT (2000)} \\

\cline{2-4}
\cline{6-10}

\colhead{} &
\colhead{2000} &
\colhead{2001} &
\colhead{2002} &
\colhead{} &
\colhead{U} &
\colhead{B} &
\colhead{V} &
\colhead{R} &
\colhead{I} 
}

\startdata 

P$_{instrumental}$\tablenotemark{a}& 
0.081{\%}& 
0.083{\%}& 
0.084{\%}& 
 &
0.029{\%}& 
0.018{\%}& 
0.023{\%}& 
0.021{\%}& 
0.069{\%} \\

$\theta - \theta _{published}$\tablenotemark{b}& 
42.7$^{o}$& 
41.8$^{o}$& 
40.4$^{o}$& 
 &
-71.3$^{o}$& 
-71.7$^{o}$& 
-71.7$^{o}$& 
-71.9$^{o}$& 
-70.6$^{o}$ \\

\enddata

\tablenotetext{a}{These values represent the average observed polarization of the assumed unpolarized stars HD 154345 (Gehrels 1974), HD 9407 (Gliese 1969) and HD 212311 (Schmidt, Elston, {\&} Lupie 1992) for our McDonald (unfiltered) observations and HD 154345 (Gehrels 1974), HD 9407 (Gliese 1969) and HD 67228 (Piirola 1977) for our NOT observations.}

\tablenotetext{b}{HD 204827 was chosen as our standard polarized star for both sites with $\theta _{published}=59.3^o$ according to Hsu {\&} Breger 1982.}

\end{deluxetable}

\clearpage

\begin{deluxetable}{lcccccc} 
\tablecolumns{8} 
\tabletypesize{\scriptsize}
\rotate
\tablewidth{0pc} 
\tablecaption{Polarimetric Parameters of CX Cep from our Unfiltered Observations} 
\tablehead
{ 
\colhead{} 			& 
\multicolumn{4}{c}{Unfiltered Results (McDonald)} &
\colhead{Schulte-Ladbeck {\&} } &
\colhead{Kartasheva} \\
\cline{2-5}

\colhead{} &
\colhead{2000} &
\colhead{2001} &
\colhead{2002} &
\colhead{Combined} &
\colhead{van der Hucht (1989)} &
\colhead{(2002b)}

\\
\colhead{} &
\colhead{} &
\colhead{} &
\colhead{} &
\colhead{} &
\colhead{Unfiltered} &
\colhead{V band}

}

\startdata

U$_{0}$ ({\%})& 
6.068 $\pm $ 0.002& 
6.176 $\pm $ 0.004& 
6.076 $\pm $ 0.007& 
6.095 $\pm $ 0.001& 
6.14 $\pm $ 0.04& 
7.04 $\pm $ 0.07 \\

U$_{1}$ ({\%})& 
-0.022 $\pm $ 0.002& 
-0.047 $\pm $ 0.007& 
0.075 $\pm $ 0.006& 
-0.013 $\pm $ 0.002& 
0.08 $\pm $ 0.06& 
-0.11 $\pm $ 0.07 \\

U$_{2}$ ({\%})& 
-0.027 $\pm $ 0.002& 
-0.003 $\pm $ 0.003& 
0.049 $\pm $ 0.010& 
-0.009 $\pm $ 0.002& 
-0.06 $\pm $ 0.06& 
-0.18 $\pm $ 0.07 \\

U$_{3}$ ({\%}) \quad & 
0.053 $\pm $ 0.002& 
0.057 $\pm $ 0.006& 
-0.023 $\pm $ 0.006& 
0.018 $\pm $ 0.002& 
0.08 $\pm $ 0.06& 
0.06 $\pm $ 0.04 \\

U$_{4}$ ({\%}) \quad & 
0.182 $\pm $ 0.003& 
0.196 $\pm $ 0.005& 
0.185 $\pm $ 0.007& 
0.188 $\pm $ 0.002& 
0.17 $\pm $ 0.06& 
0.31 $\pm $ 0.09 \\

Q$_{0}$ ({\%}) \quad & 
-0.390 $\pm $ 0.002& 
-0.400 $\pm $ 0.004& 
-0.476 $\pm $ 0.007& 
-0.413 $\pm $ 0.001& 
-0.01 $\pm $ 0.04& 
-0.08 $\pm $ 0.07 \\

Q$_{1}$ ({\%}) \quad & 
0.006 $\pm $ 0.002& 
-0.162 $\pm $ 0.007& 
-0.034 $\pm $ 0.006& 
-0.025 $\pm $ 0.002& 
-0.05 $\pm $ 0.05& 
0.13 $\pm $ 0.07 \\

Q$_{2}$ ({\%}) \quad & 
-0.006 $\pm $ 0.002& 
0.047 $\pm $ 0.003& 
-0.046 $\pm $ 0.010& 
-0.003 $\pm $ 0.002& 
-0.04 $\pm $ 0.06& 
0.04 $\pm $ 0.07 \\

Q$_{3}$ ({\%}) \quad & 
-0.240 $\pm $ 0.002& 
-0.180 $\pm $ 0.006& 
-0.153 $\pm $ 0.006& 
-0.200 $\pm $ 0.002& 
-0.12 $\pm $ 0.05& 
-0.25 $\pm $ 0.04 \\

Q$_{4}$ ({\%}) \quad & 
-0.056 $\pm $ 0.003& 
-0.119 $\pm $ 0.005& 
-0.062 $\pm $ 0.007& 
-0.076 $\pm $ 0.002& 
0.04 $\pm $ 0.05& 
-0.19 $\pm $ 0.09 \\

$i$ ($^o$)& 
73$\pm$2& 
79$\pm$3& 
63$\pm$20& 
70$\pm$2& 
68$\pm$18& 
79$\pm$12 \\

$\Omega $\tablenotemark{b} ($^o$) & 
75$\pm $1& 
69$\pm $2& 
61$\pm $6& 
72$\pm $1& 
50$\pm $38& 
67$\pm $13 \\

$\tau _{o}$G\tablenotemark{a}  ({\%})& 
0.018& 
0.087& 
0.053& 
0.015& 
0.067& 
0.127 \\

$\tau _{o}$H\tablenotemark{a}  ({\%})& 
0.250& 
0.269& 
0.165& 
0.219& 
0.166& 
0.397 \\

rms$_{U}$ ({\%})& 
0.059& 
0.071& 
0.065& 
0.082& 
0.170& 
0.150 \\

rms$_{Q}$ ({\%})& 
0.075& 
0.080& 
0.077& 
0.103& 
0.121& 
0.177 \\

\enddata

\tablenotetext{a}{The errors on $\tau _{o}$G and $\tau _{o}$H were found to be less than 0.02\% at the $2\sigma$ level and we decided not to present them explicitly.}

\tablenotetext{b}{Our definition for $\Omega$ is that of Harries {\&} Howarth (1996).}

\end{deluxetable}

\begin{deluxetable}{lccccccccccc} 
\tablecolumns{12} 
\tablewidth{0pc} 
\rotate
\tabletypesize{\scriptsize}
\tablecaption{Parameters of the Non-BME78 models of CX Cep} 
\tablehead
{ 
\colhead{} 			& 
\colhead{M1} &
\colhead{M2} &
\colhead{M3} &
\colhead{M4} &
\colhead{M5} &
\colhead{M6} &
\colhead{M7} &
\colhead{M8} &
\colhead{M9} &
\colhead{M10} &
\colhead{M11} 

}

\startdata

$i$ ($^{o})$& 
65& 
60& 
70& 
65& 
65& 
65& 
65& 
65& 
65&
65&
65 \\

$\beta $& 
5& 
5& 
5& 
4& 
6& 
5& 
5& 
5& 
5&
5&
5 \\

R$_{O}$ ($R_{\odot}$)& 
11& 
11& 
11& 
11& 
11& 
9& 
13& 
11& 
11&
11&
11 \\

R$_{s}$ ($R_{\odot}$)& 
2.0& 
2.0& 
2.0& 
2.0& 
2.0& 
2.0& 
2.0& 
1.0& 
3.0&
2.0&
2.0 \\

$\tau_{max}$ &
0.67&
0.67&
0.67&
0.67&
0.67&
0.67&
0.67&
0.67&
0.67&
0.33&
1.00 \\

$\Delta \phi $& 
0.033& 
0.033& 
0.033& 
0.033& 
0.032& 
0.032& 
0.035& 
0.033& 
0.031&
0.031&
0.033 \\

$\Delta $U ({\%})& 
4.96& 
4.96& 
4.95& 
4.96& 
4.95& 
4.96& 
4.95& 
4.97& 
4.95&
4.95&
4.95 \\

$\Delta $Q ({\%})& 
-3.74& 
-3.70& 
-3.77& 
-3.74& 
-3.75& 
-3.73& 
-3.74& 
-3.76& 
-3.72&
-3.72&
-3.73 \\

$\gamma (10^{-17} M_{\odot}/km)$& 
6.1& 
5.4& 
6.6& 
6.6& 
5.6& 
5.5& 
6.9& 
8.4& 
4.9& 
9.3& 
4.8 \\

$\dot{M} (10^{-5} M_{\odot}/yr)$\tablenotemark{a}& 
0.33& 
0.29& 
0.36& 
0.36& 
0.30& 
0.30& 
0.37& 
0.45& 
0.27&
0.50&
0.26 \\

$rms$ (\%)\tablenotemark{b} &
0.040&
0.043&
0.047&
0.042&
0.043&
0.043&
0.043&
0.051&
0.053&
0.052&
0.043\\

\enddata

\tablenotetext{a}{The mass-loss rate was determined assuming a typical terminal velocity (i.e. $V_{\infty}$) for the wind of a WN5 star of 1700 km/s.}

\tablenotetext{b}{This $rms$ represents the average scatter of our binned data points from the modeled curve calculated by excluding any points in Stokes Q located at $\pm0.1$ in phase from the eclipses .}

\end{deluxetable}

\begin{deluxetable}{lccccc} 
\tablecolumns{6} 
\tabletypesize{\scriptsize}
\rotate
\tablewidth{0pc} 
\tablecaption{Polarimetric Parameters of CX Cep from our Multi-Band Observations} 
\tablehead
{ 
\colhead{} 	& 
\colhead{U} 	& 
\colhead{B} 	& 
\colhead{V} 	& 
\colhead{R} &
\colhead{I} 
}

\startdata

U$_{0}$ ({\%})& 
5.43 $\pm $ 0.04& 
6.48 $\pm $ 0.04& 
6.81 $\pm $ 0.04& 
6.47 $\pm $ 0.03& 
5.41 $\pm $ 0.03 \\

U$_{1}$ ({\%})& 
-0.04 $\pm $ 0.02& 
0.07 $\pm $ 0.02& 
0.07 $\pm $ 0.02& 
-0.03 $\pm $ 0.01& 
-0.02 $\pm $ 0.01 \\

U$_{2}$ ({\%})& 
-0.12 $\pm $ 0.04& 
0.09 $\pm $ 0.04& 
0.06 $\pm $ 0.05& 
0.11 $\pm $ 0.03& 
-0.18 $\pm $ 0.03 \\

U$_{3}$ ({\%}) \quad & 
0.19 $\pm $ 0.05& 
0.08 $\pm $ 0.04& 
-0.07 $\pm $ 0.05& 
0.11 $\pm $ 0.03& 
0.23 $\pm $ 0.04 \\

U$_{4}$ ({\%}) \quad & 
0.17 $\pm $ 0.03& 
0.19 $\pm $ 0.03& 
0.23 $\pm $ 0.04& 
0.09 $\pm $ 0.02& 
0.02 $\pm $ 0.02 \\

Q$_{0}$ ({\%}) \quad & 
-0.29 $\pm $ 0.04& 
-0.18 $\pm $ 0.03& 
-0.42 $\pm $ 0.03& 
-0.47 $\pm $ 0.03& 
-0.75 $\pm $ 0.03 \\

Q$_{1}$ ({\%}) \quad & 
0.03 $\pm $ 0.02& 
-0.02 $\pm $ 0.02& 
-0.01 $\pm $ 0.01& 
-0.09 $\pm $ 0.01& 
-0.05 $\pm $ 0.01 \\

Q$_{2}$ ({\%}) \quad & 
0.16 $\pm $ 0.04& 
-0.07 $\pm $ 0.04& 
0.09 $\pm $ 0.04& 
0.01 $\pm $ 0.03& 
0.03 $\pm $ 0.03 \\

Q$_{3}$ ({\%}) \quad & 
-0.18 $\pm $ 0.04& 
-0.38 $\pm $ 0.03& 
0.00 $\pm $ 0.04& 
-0.05 $\pm $ 0.03& 
-0.11 $\pm $ 0.03 \\

Q$_{4}$ ({\%}) \quad & 
-0.18 $\pm $ 0.03& 
-0.02 $\pm $ 0.03& 
-0.20 $\pm $ 0.03& 
-0.21 $\pm $ 0.02& 
-0.04 $\pm $ 0.02 \\

$i$ ($^o$)& 
89$\pm $7& 
76$\pm $14& 
86$\pm $9& 
81$\pm $8& 
87$\pm $8 \\

$\Omega$\tablenotemark{b} ($^o$) & 
67$\pm $7& 
82$\pm $8& 
65$\pm $7& 
74$\pm $7& 
58$\pm $8\\

$\tau _{o}$G\tablenotemark{a}  ({\%})& 
0.10& 
0.07& 
0.07& 
0.08& 
0.09 \\

$\tau _{o}$H\tablenotemark{a} ({\%})& 
0.36& 
0.37& 
0.31& 
0.24& 
0.26 \\

rms$_{U}$ ({\%})& 
0.16& 
0.18& 
0.17& 
0.09& 
0.15 \\

rms$_{Q}$ ({\%})& 
0.18& 
0.19& 
0.18& 
0.06& 
0.24 \\

\enddata

\tablenotetext{a}{The errors on $\tau _{o}$G and $\tau _{o}$H were found to be less than 0.02\% at the $2\sigma$ level and we decided not to present them explicitly.}

\tablenotetext{b}{Our definition for $\Omega$ is that of Harries {\&} Howarth (1996).}

\end{deluxetable}

\begin{deluxetable}{lccc} 
\tablecolumns{4} 
\tablewidth{0pc} 
\tabletypesize{\footnotesize}
\tablecaption{Interstellar and Intrinsic Components of the Polarization from CX Cep} 
\tablehead
{ 
\colhead{} 			& 
\colhead{Model 1} &
\colhead{Model 2} &
\colhead{Model 3}
}

\startdata 

P$_{max}$ ({\%})& 
10.17& 
6.75 &
6.09
\\

$\lambda _{max}$ (nm)& 
545& 
557 &
554
\\

$C_{1}$ ($^{o})$& 
41& 
51 &
46
\\

C$_{2}$ ($^{o}$nm)& 
-31& 
-52 &
-58
\\

U$_{WR}$ ({\%})& 
-3.05& 
0\tablenotemark{*} &
0.60
\\

Q$_{WR}$ ({\%})& 
-2.36& 
0\tablenotemark{*} &
-1.03
\\

P$_{WR}$ ({\%})& 
3.89& 
0\tablenotemark{*} &
1.19
\\

$\theta _{WR}$ ($^{o})$& 
26& 
N/A &
165\tablenotemark{*}
\\

rms ({\%}) & 
0.09& 
0.22 &
0.24
\\

\enddata
\tablenotetext{*}{Fixed.}

\end{deluxetable}

%% The following command ends your manuscript. LaTeX will ignore any text

\end{document}